\documentclass[preprint,12pt]{elsarticle}

\usepackage{amssymb,amsmath,amsthm,bm}
\usepackage{ dsfont }
\usepackage{color}

\newcommand{\Cov} {\mbox{$\rm{Cov}$\,}}
\newcommand{\Var} {\mbox{$\rm{Var}$\,}}

\usepackage{lineno}

\journal{Theoretical Population Biology}

\begin{document}

\begin{frontmatter}



\title{Mutation in Populations Governed by a Galton-Watson Branching Process}

\author[label1,label2]{Conrad J.\ Burden}
\ead{conrad.burden@anu.edu.au}
\author[label1]{Yi Wei}
\ead{u5166590@anu.edu.au}
\address[label1]{Mathematical Sciences Institute, Australian National University, Canberra, Australia}
\address[label2]{Research School of Biology, Australian National University, Canberra, Australia}

\begin{abstract}
A population genetics model based on a multitype branching process, or equivalently a Galton-Watson branching process for multiple alleles, is presented.  
The diffusion limit forward Kolmogorov equation is derived for the case of neutral mutations.  The asymptotic stationary solution is obtained and has the property that 
the extant population partitions into subpopulations whose relative sizes are determined by mutation rates.  An approximate time-dependent solution is obtained in 
the limit of low mutation rates.  This solution has the property that the system undergoes a rapid transition from a 
drift-dominated phase to a mutation-dominated phase 
in which the distribution collapses onto the asymptotic stationary distribution.  The changeover point of the transition is determined by the per-generation 
growth factor and mutation rate.  The approximate solution is confirmed using numerical simulations.  
\end{abstract}


\end{frontmatter}



\section{Introduction}
\label{sec:Introduction}

Since their introduction to the field by \citet{haldane1927mathematical}, Galton-Watson (GW) branching processes have 
been an important part of the population genetics landscape~\citep{patwa2008fixation}.  For example, probabilities of non-extinction derived through 
branching process approximations play an indispensable role in many complex population models~\citep[e.g.][]{desai2007beneficial}.  However, 
as argued by \citet{{mode2013inclusion}}, 
the influence of models based on GW branching processes has in general  been overshadowed, at least within the text book literature, 
by that of Wright-Fisher (WF) based models.  Much of the WF model's dominance can be attributed to the intuitive appeal of the 
coalescent~\citep{kingman1982coalescent}, which is a natural consequence of WF models but mathematically 
formidable for a GW process~\citep{lambert2013coalescent}, and to the WF model's well-known diffusion limit 
via the forward Kolmogorov equation, as championed by \citet{Kimura:1955fk,kimura1955stochastic,kimura1964diffusion}.  

Somewhat lesser known than the work of Kimura, and predating it by four years, 
is a solution to the diffusion limit of a GW branching process published by~\citet{feller1951diffusion}.  
It is surprising that, although Feller's solution was presented in the context of genetics, the vast majority of applications of 
Feller's solution have been to areas other than genetics \citep[see][and references therein]{gan2015singular}.  It is equally surprising that when 
population genetics per se is modelled as a branching process, it is generally as a discrete state space simulation~\citep{mode2012stochastic,cyran2010alternatives} 
or a continuous birth-death process~\citep{stadler2015well}, without reference to Feller's diffusion limit.  

This paper follows on from an earlier work~\citep{burden2016genetic} in which Feller's diffusion limit is exploited to study genetic drift in haploid populations 
governed by a GW branching process.  In that work it was shown that, in the absence of mutations and selection, expected fixation times and 
probabilities of fixation for a critical branching process match those of the WF model.  However, for a supercritical branching process there is 
a finite probability that an allele will  never fix.  The dynamics of the branching process enabled an estimate to be made of the time since the most recent 
common ancestor of an extant population, for instance, mitochondrial Eve.  

The current paper extends the branching model to a multi-allelic population with mutations, and is equivalent to a multitype branching 
process~\citep{mode1971multitype,haccou2005branching}.  Multitype branching processes have been applied in population science to modelling 
cancers~\citep{durrett2010evolution,iwasa2003evolutionary}, modelling bacterial cultures~\citep{wahl2015survival}, and in ecological 
modelling~ \cite[][Chapter~15]{antia2003role,caswell2001matrix}.

Our model is set out in detail in Section~\ref{sec:TheModel}, and the diffusion limit forward Komogorov equation is derived in 
Section~\ref{sec:DiffusionLimit}.  Our choice of diffusion limit is such that continuum time is scaled by the log of the per-generation growth factor $\lambda$, 
and the population size is scaled by the mean exponential growth.  This leads to a slightly more elegant forward Kolmogorov equation than Feller's original, 
but with the same physical interpretation (see Eq.~(\ref{1AlleleKolmogEq})).  Our scaling has the disadvantage that it is not suitable for critical growth, $\lambda = 1$, 
thus limiting our analysis to the supercritical case.  On the other hand it has the advantage that the solution is classified in terms of a 
1-parameter family of density functions (see Eq.~(\ref{fellerDistrDef}) and (\ref{f1AlleleAndFeller})).   

In Section~\ref{sec:MutationZero} Feller's method of solution via a Laplace transform for the 1-allele case is briefly summarised in order to 
facilitate analysis of the case of non-zero mutations in Sections~\ref{sec:MutationNonZero} and \ref{sec:2Alleles}.  Although we are unable to find a complete 
analytic solution, we are able to obtain the asymptotic stationary solution for the case of 2 alleles, and also an approximate solution for all times in the biologically 
realistic limit of low mutation rates.  An interesting result is that the solution undergoes a rapid changeover in behaviour from a perturbation on the 
zero-mutation solution to an asymptotic collapse onto a state in which any extant population partitions into subpopulations in proportions determined by 
mutation rates.  Section~\ref{sec:NumericalSimulations} is devoted to numerical simulations to confirm our analytical results, and to confirm consistency of the model with 
mitochondrial genomic data.  Section~\ref{sec:Discussion} is devoted to a discussion and conclusions.  


\section{The model}
\label{sec:TheModel}

We consider a population of $M(t)$ haploid individuals which are assumed to reproduce in discrete, non-overlapping 
generations $t = 0, 1, 2, \ldots$.  The population is subdivided at any generation into $K$ allele types, and the number of 
copies of type $i$ within the population is $Y_i(t)$.  Thus 
\begin{equation}
\sum_{i = 1}^K Y_i(t) = M(t).  \label{Mdef}
\end{equation}
The individuals are assumed to reproduce according to a GW process whereby the number of 
offspring per individual of allele type $i$ is given by a set of identically and independently distributed (i.i.d.) random variables $S_\alpha^{(i)}$, 
$\alpha = 1, \ldots, Y_i(t)$, whose common distribution is denoted by a generic non-negative integer 
valued random variable $S^{(i)}$ with mean and variance 
\begin{equation}
E(S^{(i)}) = \lambda_i, \qquad \Var(S^{(i)}) = \sigma_i^2, \qquad i = 1, \ldots, K.  \label{lambdaSigmaDef}
\end{equation}

\begin{figure}[t!]
\begin{center}
\centerline{\includegraphics[width=0.9\textwidth]{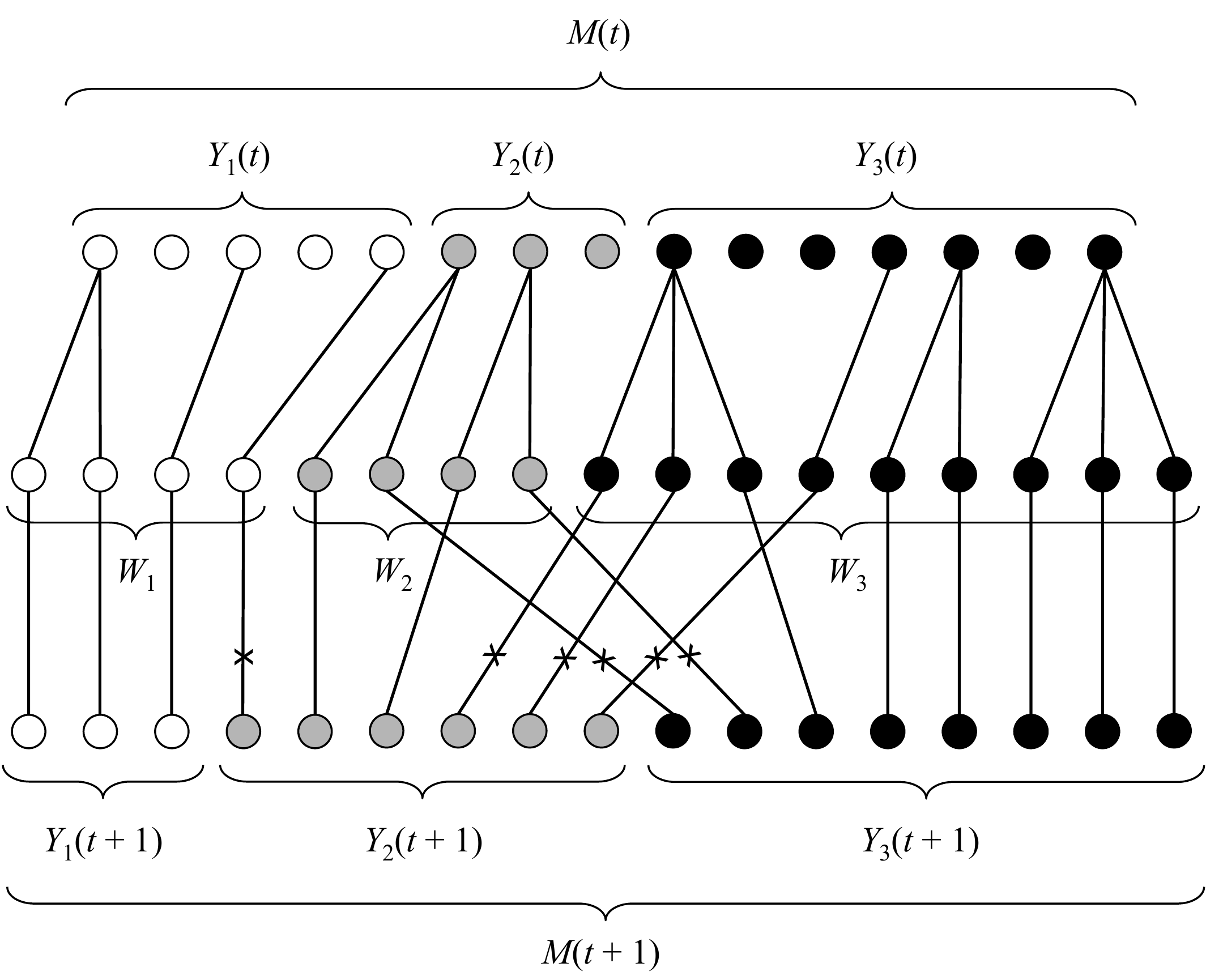}}
\caption{One time step of the GW model with mutations: At time step $t$ a population of $M(t)$ individuals is partitioned into 
subsets containing $Y_i(t)$ individuals of allele type $i$.  Each individual generates a random number of offspring of the same allele 
type as its parent, and the number of offspring initially of type $i$ is defined as $W_i$.  Individuals may mutate during their lifetime to create the 
new generation containing $Y_i(t + 1)$ individuals of allele type $i$.  Timelines of individuals who have changed their identity during maturation are 
marked with a $\times$.} 
\label{fig:OneTimestep}
\end{center}
\end{figure}

Furthermore the alleles are assumed to 
undergo random mutations from type $i$ to type $j$ at a rate $u_{ij}$ per individual per generation, where 
\begin{equation}
u_{ij} \ge 0, \qquad \sum_{j = 1}^K u_{ij} = 1.  \label{uProperties}
\end{equation}
A single time step is illustrated in Fig~\ref{fig:OneTimestep}.  

Define the number offspring born to parents of allele type $i$ in generation $t$ to be 
\begin{equation}
W_i = \sum_{\alpha = 1}^{Y_i(t)} S_\alpha^{(i)}.  \label{WDef}
\end{equation}
During its lifetime the new generation undergoes mutations, culminating in a new mature generation in which the number of 
individuals of type $i$ 
is expressible as a sum of random variables
\footnote{Throughout the paper, a vector of length $K$ will be denoted in bold type, e.g.\ ${\bf W} = (W_1, \ldots, W_K)$.}  
\begin{equation}
Y_i(t + 1)|{\bf W} = V_{1i} + V_{2i} + \ldots V_{Ki}, \label{VjiDef}
\end{equation}
where $V_{ji}$ is the number of individuals who begin life as allele type $j$ and mature to become allele type $i$.  For fixed parental type $j$ 
the $V_{ji}$ have a multinomial distribution: 
\begin{equation}
(V_{j1}, \ldots, V_{jK}) \sim  \text{Multinom}(W_j, (u_{j1}, \ldots, u_{jK})).  \label{VjiDistrib}
\end{equation}
Note also that for fixed $i$ and conditional on ${\bf Y}(t)$, the $V_{ji}$ are independent.  

In the following we make use of the convention that, given two random variables $X_1$ and $X_2$, $E(X_1 | X_2)$ and $\Var(X_1 | X_2)$ 
represent the random variables $g(X_2)$ and $h(X_2)$ respectively, where $g(x) = E(X_1 | X_2 = x)$ and $h(x) = \Var(X_1 | X_2 = x)$
\citep[see][Def.~3.7.3]{grimmett2001probability}.  
From Eqs.~(\ref{lambdaSigmaDef}) and (\ref{WDef}) and the independence of the $W_i|{\bf Y}(t)$ we have that 
\begin{equation}
\begin{split}
E(W_i|{\bf Y}(t)) &= \lambda_i Y_i(t), \\
\Var(W_i|{\bf Y}(t)) &= \sigma_i^2 Y_i(t), \\
\Cov(W_i, W_j|{\bf Y}(t)) &= 0, \qquad \text{for }i \ne j, 
\end{split} \label{ExpAndVarW}
\end{equation}
while from Eqs.~(\ref{VjiDef}) and (\ref{VjiDistrib}) we have that 
\begin{equation}
\begin{split}
E(Y_i(t + 1)|{\bf W}) & = \sum_{j = 1}^K u_{ji}W_j, \\
\Var(Y_i(t + 1)|{\bf W}) & = \sum_{j = 1}^K u_{ji}(1 - u_{ji})W_j, \\
\Cov(Y_i(t + 1), Y_j(t + 1)|{\bf W}) & = -\sum_{k = 1}^K u_{ki}u_{kj}W_k,  \qquad \text{for }i \ne j.
\end{split} \label{ExpAndVarYtPlus1}
\end{equation}
Recall the laws of total expectation, total variance and total covariance which state that for any random variables $A$, $B$ and $C$, 
\begin{equation}
\begin{split}
E(A) &= E(E(A|B)), \\
\Var(A) &= E(\Var(A|B)) + \Var(E(A|B)), \\
\Cov(A|B) &= E(\Cov(A, B|C)) + \Cov(E(A|C), E(B|C)). 
\end{split}	\label{totalLaws}
\end{equation}
Applying these laws to Eqs.~(\ref{ExpAndVarW}) and (\ref{ExpAndVarYtPlus1}) one obtains 
\begin{equation}
\begin{split}
E(Y_i(t + 1)|{\bf Y}(t)) &= \sum_{j = 1}^K \lambda_j u_{ji} Y_j(t) \\
\Var(Y_i(t + 1)|{\bf Y}(t)) &= \sum_{j = 1}^K \{\lambda_j u_{ji}(1 - u_{ji}) + \sigma_j^2 u_{ji}^2 \} Y_j(t) \\
\Cov(Y_i(t + 1), Y_j(t + 1)|{\bf Y}(t)) &= \sum_{k = 1}^K (\sigma_k^2 - \lambda_k) u_{ki}u_{kj}  Y_k(t),  \qquad \text{for }i \ne j. 
\end{split}		\label{YtPlus1GivenYt}
\end{equation} 

The scenario described above is an example of a {\em multitype branching process}~\citep{mode1971multitype,haccou2005branching}, 
for which various limit theorems have been proven.  More specifically, suppose we define a $K \times K$ matrix $\mu$ whose $(ij)^\text{th}$ element is 
the expected number of offspring of type-$j$ from a parent of type-$i$.  In our case 
\begin{equation}
\mu_{ij} = \lambda_i u_{ij}. \label{muDef}
\end{equation}
Since all its elements are non-negative, $\mu$ has a unique positive real left eigenvalue $\rho$, say, which is larger in absolute value than any other 
left eigenvalue.  If the corresponding eigenvector is $\pmb{\nu}$, and $\rho > 1$, then it can be shown 
that~\citep[see][Section~1.8 and references therein]{mode1971multitype} 
\begin{equation}
\lim_{t \rightarrow \infty} \rho^{-t} {\bf Y}(t) = X \pmb{\nu}, 	\label{limitTheorem}
\end{equation}
almost surely, where the distribution of the random variable $X$ depends on the distribution of ${\bf Y}(0)$.  The continuum limit of this result 
will manifest in Section~\ref{sec:Stationary} for the $K = 2$ case.  

As it stands the model encapsulated in Eq.~(\ref{YtPlus1GivenYt}) includes not only mutations, but also selection: 
Those alleles with with higher values of $\lambda_i$ will produce 
more offspring on average and therefore be selected for, while those with lower $\lambda_i$ will be selected against.  For the remainder of 
the paper we will consider only neutral mutations in a growing population.  That is, from here on we assume the $S_\alpha^{(i)}$ in 
Eq.~(\ref{WDef}) are i.i.d.\ across all allele types, and represented by a common random variable $S$, independent of $i$.  
Accordingly we set all $\lambda_i$ to a common value $\lambda$ and all $\sigma_i^2$ to a common value $\sigma^2$ in Eq.~(\ref{YtPlus1GivenYt}).  
Note that with this assumption the total number of offspring of 
parents alive at time step $t$ is, from Eq.~(\ref{WDef}), $\sum_{i = 1}^{K} W_k = \sum_{\alpha = 1}^{M(t)} S_\alpha$.  Since the mutation step 
in Fig.~\ref{fig:OneTimestep} does not change the total population size we therefore have that 
\begin{equation}
M(t + 1) = \sum_{\alpha = 1}^{M(t)} S_\alpha, 
\end{equation} 
and so for neutral evolution, the total population size $M(t)$ is effectively a 1-allele GW process.  


\section{Diffusion limit of neutral evolution}
\label{sec:DiffusionLimit}

The diffusion limit of the above model was studied in the absence of mutations (i.e.\ with the $u_{ij} = 0$) by~\citet{burden2016genetic}.  
We set the initial conditions as 
\begin{equation}
M(0) = m_0, \qquad Y_i(0) = z_{0i} m_0, \quad i = 1, \ldots K, \label{initCond}
\end{equation}
where $z_{0i} \ge 0$ are initial relative allele frequncies satisfying $\sum_{i = 1}^K z_{0i} = 1$.  \citet{burden2016genetic} defined the 
the diffusion limit as the limit $m_0 \rightarrow \infty$, $\lambda \rightarrow 1$, taken in such a way that $\sigma^2$, $z_{0i}$, and the 
product $m_0 \log \lambda$ are held fixed\footnote{If, on the other hand, one takes the limit $\lambda \rightarrow 1$, $\sigma^2 \rightarrow 0$ such that 
$m_0$ and $\sigma^2/\log\lambda$ remain fixed, a birth-death process is obtained \citep[][p165]{Cox78}.  
We believe the diffusion limit to be more appropriate to population genetics 
and in particular to comparison with conventional WF dynamics than a birth-death process~\citep{stadler2015well}, which is more relevant to 
phylogenetic trees.}.  
In particular, it was found that provided the growth rate $\lambda$ is close to but not equal to 1, the dynamics is 
entirely determined by the parameter 
\begin{equation} 
\kappa_0 = \frac{2 m_0 \log \lambda}{\sigma^2},	\label{kappa0Def}
\end{equation}
and the initial allele abundances $z_{0i}$.  In the absence of mutations, 
for the supercritical case $\lambda > 1$ 
the forward Kolmogorov (or Fokker-Planck) equation takes a particularly elegant form (see Eq.~(\ref{1AlleleKolmogEq}) below) provided 
a continuous time $s$ and an infinitesimal time step $\delta s$
are defined as~\citep[see][Section~4]{burden2016genetic}  
\begin{equation}
s = t \log \lambda, \qquad \delta s = \log \lambda, \label{sDef}
\end{equation}
while defining exponentially rescaled allele abundances and total population size 
\begin{equation}
Z_i(s) = \frac{1}{m_0 \lambda^t} Y_i(t), \qquad 
Z_{\rm tot}(s) = \frac{1}{m_0 \lambda^t} M(t) = \sum_{i = 1}^K Z_i(s),  \label{ZDef}
\end{equation}
and initial conditions 
\begin{equation}
Z_i(0) = z_{0i}, \qquad Z_\text{tot}(0) = 1. 		\label{ZInitConditions}
\end{equation}
One can readily check that, because $M(t)$ is a GW process, 
\begin{equation}
E(Z_{\rm tot}(s)) = \sum_{i = 1}^K E(Z_i(s)) = 1,  
\end{equation}
even in the presence of mutations.  

In order to include mutations, we must also introduce per-unit-continuous-time mutation rates $r_{ij}$ which we define as 
\begin{equation} 
u_{ij} = \begin{cases}
r_{ij} \,\delta s & \text{if } i \ne j; \\
1 - \sum_{\{k: k \ne i\}} r_{ik} \,\delta s    & \text{if } i = j, \label{rFromU} 
\end{cases}
\end{equation}
or equivalently
\begin{equation}
r_{ij} = \frac{u_{ij}}{\log\lambda} \qquad \text{if } i \ne j, 	\label{rijDef}
\end{equation}
on the understanding that $u_{ij} \rightarrow 0$ in the continuum limit for $i \ne j$.  

In general, a multi-dimensional forward Kolmogorov equation for the time dependent joint density $f_{\bf Z}({\bf z}, s)$ of the continuous random variables 
${\bf Z} = (Z_i, \ldots, Z_K)$ takes the form~\citep[][Section~4.8]{Ewens:2004kx}  
\begin{equation} 
\frac{\partial f_{\bf Z}({\bf z}, s)}{\partial s} = - \sum_{i = 1}^K \frac{\partial}{\partial z_i} \left\{ a_i({\bf z}) f_{\bf Z}({\bf z}, s) \right\} 
		+ \frac{1}{2} \sum_{i, j = 1}^K \frac{\partial^2}{\partial z_i \partial z_j} \left\{ b_{ij}({\bf z}) f_{\bf Z}({\bf z}, s) \right\}, \label{genericKolmogEq}
\end{equation}	
where the functions $a_i({\bf z})$ and $b_{ij}({\bf z})$ are determined by the incremental expectations 
\begin{equation}
\begin{split}
E(\delta Z_i(s) | {\bf Z}(s) &= {\bf z}) = a_i({\bf z}) \delta s + o(\delta s), \\
E(\delta Z_i(s) \delta Z_j(s) | {\bf Z}(s) &= {\bf z}) = b_{ij}({\bf z}) \delta s + o(\delta s), 
\end{split}
\end{equation}
where $\delta Z_i(s) = Z_i(s + \delta s) - Z_i(s)$.   Combining the definitions Eqs.~(\ref{kappa0Def}), (\ref{sDef}), (\ref{ZDef}) and (\ref{rFromU}), 
with the expectation values Eq.~(\ref{YtPlus1GivenYt}) one obtains, after some working, the forward Kolmogorov equation for our model: 
\begin{equation}
\frac{\partial  f_{\bf Z}({\bf z}, s; {\bf z}_0)}{\partial s} = \sum_{i, j = 1}^K \frac{\partial}{\partial z_i} \left\{ (r_{ij} z_i - r_{ji} z_j) f_{\bf Z}({\bf z}, s; {\bf z}_0) \right\} 
	 + \frac{e^{-s}}{\kappa_0} \sum_{i = 1}^K \frac{\partial^2}{\partial z_i^2} \left\{z_i f_{\bf Z}({\bf z}, s; {\bf z}_0) \right\}.    \label{KAlleleKolmogEq}
\end{equation}
Note that we have introduced the notation of including the initial allele frequencies in the argument of the function $f_{\bf Z}$ to indicate the solution 
corresponding to the initial conditions 
\begin{equation}
f_{\bf Z}({\bf z}, 0; {\bf z}_0) = \delta({\bf z} - {\bf z}_0) = \prod_{i = 1}^K \delta(z_i - z_{0i}). \label{KAlleleInitCond}
\end{equation} 
We have been unable to find a full analytic solution to Eqs.~(\ref{KAlleleKolmogEq}) and (\ref{KAlleleInitCond}).  However, we are able to characterise 
the solution in various limiting cases.  It will prove instructive to begin with the situation in which the mutation rates are set to zero.  


\section{Mutation rates set to zero}
\label{sec:MutationZero}

If the mutation rates $r_{ij}$ are set to zero in Eq.~(\ref{KAlleleKolmogEq}) the system decouples into a set of independent 
random variables $Z_i$.  In particular the joint distribution takes the form 
\begin{equation}
f_{\bf Z}({\bf z}, s; {\bf z}_0) = \prod_{i = 1}^K f_{Z_i}(z_i, s; z_{0i}, \kappa_0), \label{zeroMutationDensity}
\end{equation}
where each individual $f_{Z_i}(z_i, s; z_{0i}, \kappa_0)$ evolves according to the following forward Kolmogorov equation describing an independent 1-allele GW process: 
\begin{equation}
\frac{\partial f_\text{1-allele}(z, s; z_0, \kappa_0)}{\partial s} = \frac{e^{-s}}{\kappa_0} \frac{\partial^2}{\partial z^2} \left(z f_\text{1-allele}(z, s; z_0, \kappa_0) \right), \label{1AlleleKolmogEq}
\end{equation}
with initial condition
\begin{equation}
f_\text{1-allele}(z, 0; z_0, \kappa_0) = \delta(z - z_0).  \label{1AlleleInitCond}
\end{equation}

\begin{figure}[t!]
\begin{center}
\centerline{\includegraphics[width=0.6\textwidth]{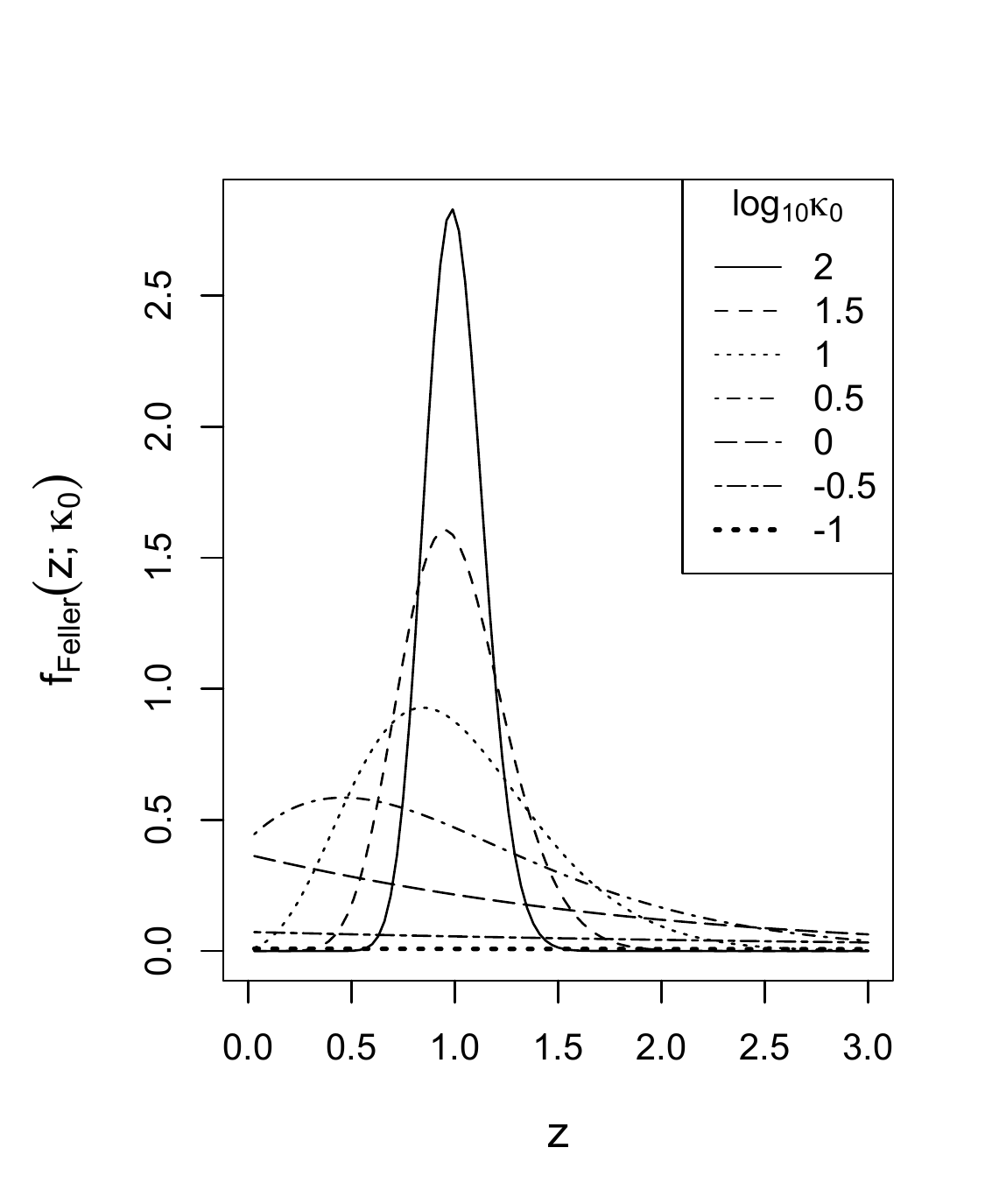}}
\caption{Plots of the continuous part of the function $f_\text{Feller}(z; \kappa_0)$, Eq.~(\ref{fellerDistrDef}), for a range of values of 
the parameter $\kappa_0$ defined by Eq.~(\ref{kappa0Def}) between $10^{-1}$ and $10^2$.  
The point mass at $z = 0$ is not shown.} 
\label{fig:FellerFunction}
\end{center}
\end{figure}

As will be verified below, the solution to Eq.~(\ref{1AlleleKolmogEq}) is 
\begin{eqnarray}
\lefteqn{f_\text{1-allele}(z, s; z_0, \kappa_0) = } \nonumber \\
 & & \delta(z)e^{-\tilde{\kappa}_0(s) z_0} + 
		\tilde{\kappa}_0(s) \left(\frac{z_0}{z}\right)^{\frac{1}{2}} e^{-\tilde{\kappa}_0(s) (z_0 + z)} I_1\left(2 \tilde{\kappa}_0(s) (z_0 z)^\frac{1}{2} \right), \label{1AlleleSoln}  
\end{eqnarray}
where 
\begin{equation}
\tilde{\kappa}_0(s) = \frac{\kappa_0}{1 - e^{-s}},   \label{kappaTildeDef}
\end{equation}
and $I_1$ is the modified Bessel function of order 1.  In the first term the coefficient of the delta-function gives the probability of the population becoming extinct up to time $s$.  
This solution is quoted in \citet{burden2016genetic} and is equivalent, up to differing notation, to a solution initially found by~\citet{feller1951diffusion,feller1951two} 
and described in~\citet[][Section~14.5]{Bailey:1964fk}.  
It can be written in terms of a 1-parameter family of density functions which we will denote by
\begin{equation}
f_\text{Feller}(z; \kappa_0) = \delta(z) e^{-\kappa_0} + \kappa_0 z^{-\frac{1}{2}} e^{-\kappa_0(1 + z)} I_1\left(2\kappa_0z^{-\frac{1}{2}}\right),   \label{fellerDistrDef}
\end{equation}
as 
\begin{equation}
f_\text{1-allele}(z, s; z_0, \kappa_0) = \frac{1}{z_0} f_\text{Feller}\left(\frac{z}{z_0}; \frac{\kappa_0 z_0}{1 - e^{-s}} \right).	\label{f1AlleleAndFeller}
\end{equation}
Plots of the continuous part of $f_\text{Feller}$ are shown in Fig.~\ref{fig:FellerFunction}.  

In order to facilitate subsequent discussion of the full model with mutations we next give a summary of the derivation of this solution, 
following a method described in~\citet[][pages~235 and 250]{Cox78}.  

\subsection{Derivation of the 1-allele solution Eq.~(\ref{1AlleleSoln})}

We define the following Laplace transform of the function $f_\text{1-allele}$: 
\begin{eqnarray} 
\phi_\text{1-allele}(\theta, s; z_0) & = & E\left( e^{-\kappa_0 \theta Z(s)}\right | Z(0) = z_0) \nonumber \\
	& = &\int_{-\infty}^\infty e^{-\kappa_0 \theta z}  f_\text{1-allele}(z, s; z_0, \kappa_0) \,dz,  \label{laplace1Dim}
\end{eqnarray}
where convergence of the integral at $-\infty$ is achieved by defining $f_\text{1-allele}(z, s; z_0, \kappa_0)$ to be zero for $z < 0$.  Applying the Laplace transform to 
both sides of Eq.~(\ref{1AlleleKolmogEq}) and carrying through straightforward manipulations gives the corresponding Bartlett's equation 
\begin{equation}
\frac{\partial \phi_\text{1-allele}(\theta, s; z_0)}{\partial s} + e^{-s} \theta^2 \frac{\partial \phi_\text{1-allele}(\theta, s; z_0)}{\partial \theta} = 0.  \label{Bartlett1Allele}
\end{equation}
The initial condition corresponding to Eq.~(\ref{1AlleleInitCond}) is 
\begin{equation} 
\phi_\text{1-allele}(\theta, 0; z_0) = e^{-\kappa_0 z_0 \theta}.  \label{Bartlett1AlleleInitCond}
\end{equation}

The Laplace transform has reduced the problem to a first-order partial differential equation which can be solved by observing that $\phi_\text{1-allele}(\theta, s; z_0)$ 
is constant along characteristic curves in the $s$-$\theta$ plane defined by 
\begin{equation}
0 = \frac{\partial \phi_\text{1-allele}(\theta, s; z_0)}{\partial s} + \frac{d \theta}{ds} \frac{\partial \phi_\text{1-allele}(\theta, s; z_0)}{\partial \theta}.  \label{characteristicDef}
\end{equation}
Comparing Eqs.~(\ref{Bartlett1Allele}) and (\ref{characteristicDef}) we see that the characteristic curves are the solutions to the differential equation 
\begin{equation}
\frac{d\theta}{ds} = e^{-s} \theta^2, 
\end{equation}
namely 
\begin{equation}
\theta(s) = \frac{\theta(0)}{1 - (1 - e^{-s})\theta(0)}. \label{characteristic1Allele}
\end{equation}
Thus, given a point $(s, \theta)$, the value of $\phi_\text{1-allele}$ is equal to its value at the point on the boundary of the $s$-$\theta$ plane obtained 
by tracing the characteristic curve back to the point 
\begin{equation}
\theta(0) = \frac{\theta}{1 + (1 - e^{-s})\theta},  
\end{equation}
obtained by inverting Eq.~(\ref{characteristic1Allele}).  Putting this together with the initial condition Eq.~(\ref{Bartlett1AlleleInitCond}) gives 
\begin{equation}
\phi_\text{1-allele}(\theta, s; z_0) = \exp\left\{ - \frac{\kappa_0 z_0\theta}{1 + (1 - e^{-s})\theta} \right\}.	\label{Bartlett1AlleleSoln}
\end{equation}
The procedure for inverting the Laplace transform of a function of the form $\phi(\theta) = \exp\{ - A\theta / 1 + B\theta) \})$ for arbitrary coefficients $A$ and $B$ is 
described in detail in~\citet[][pages~236 and and 250]{Cox78}, and leads directly to Feller's solution, Eq.~(\ref{1AlleleSoln}).  


\section{Non-zero mutation rates}
\label{sec:MutationNonZero}

We now return to the full model of neutral evolution, Eq.~(\ref{KAlleleKolmogEq}), for which, as we have remarked, a full analytic solution 
remains intractable.  Before moving on, we address two properties of the solution.  

Firstly, note that although $f_{\bf Z}({\bf z}, s; {\bf z}_0)$ is unknown, we do know that $Z_{\rm tot}(s) = \sum_{i = 1}^K Z_i(s)$ is the continuum limit of a 1-allele 
GW process, so the corresponding marginal distribution must be the 1-allele solution Eq.~(\ref{1AlleleSoln}) with $z_0$ set to 1, namely,  
\begin{equation}
f_{Z_{\rm tot}}(z, s) = f_\text{1-allele}(z, s; 1, \kappa_0).  \label{ZTotDistr}
\end{equation}
In particular, we know that the probability of extinction of the entire population up to time $s$ is $\exp(-\tilde\kappa_0(s))$.  

Secondly, we are able to obtain a Bartlett's equation for the problem, and hence in principle at least, write the solution in terms of solutions to a set of 
coupled characteristic equations.  
Define a $K$-dimensional Laplace transform of the density 
function $f_{\bf Z}({\bf z}, s; {\bf z}_0)$ by 
\begin{eqnarray}
\phi(\pmb{\theta}, s; {\bf z}_0) & = & E\left( e^{-\kappa_0 \,\pmb{\theta}.{\bf Z}(s)}\right | {\bf Z}(0) = {\bf z}_0) \nonumber \\
	& = &\int_{\mathds{R}^K} d^K{\bf z} \,e^{-\kappa_0 \,\pmb{\theta}.{\bf z}} f_{\bf Z}({\bf z}, s; {\bf z}_0),     \label{laplaceKDim}
\end{eqnarray}
where $\pmb{\theta} = (\theta_1, \ldots, \theta_K)$ and $\pmb{\theta}.{\bf z} = \sum_{i = 1}^K \theta_i z_i$.  Bartlett's equation corresponding to Eq.~(\ref{KAlleleKolmogEq}) is, after a little algebra,  
\begin{equation}
\frac{\partial \phi(\pmb{\theta}, s; {\bf z}_0)}{\partial s} + 
	\sum_{i = 1}^K \left\{ \sum_{j = 1}^K (\theta_i - \theta_j) r_{ij} + e^{-s} \theta_i^2 \right\} \frac{\partial \phi(\pmb{\theta}, s; {\bf z}_0)}{\partial \theta_i} = 0,   \label{BartlettKAllele}
\end{equation}
with initial boundary condition following from Eq.~(\ref{KAlleleInitCond})
\begin{equation}
\phi(\pmb{\theta}, 0; {\bf z}_0) = e^{-\kappa_0 \,\pmb{\theta}.{\bf z}_0}.  \label{KAlleleBartlettInitCond}
\end{equation}

By analogy with the 1-allele case we can attempt to find a solution in terms of characteristic curves in the $(K + 1)$-dimensional space spanned 
by the coordinates $(\pmb{\theta}, s)$, with the defining property 
\begin{equation}
\frac{\partial \phi(\pmb{\theta}, s; {\bf z}_0)}{\partial s} + 
	\sum_{i = 1}^K \frac{d \theta_i}{ds} \frac{\partial \phi(\pmb{\theta}, s; {\bf z}_0)}{\partial \theta_i} = 0. 
\end{equation}
These characteristics are solutions to the set of coupled ordinary differential equations 
\begin{equation}
\frac{d \theta_i}{ds} = \sum_{j = 1}^K (\theta_i - \theta_j) r_{ij} + e^{-s} \theta_i^2,  \qquad i = 1, \ldots K. \label{characteristicKAlleles}
\end{equation}
Unlike the 1-allele case, however, these equations are not separable, and no easy solution is apparent.  
For the remainder of this paper we will consider the case of $K = 2$ alleles, derive analytic solutions in certain limiting cases, and compare the 
behaviour of these solutions to numerical simulations.  


\section{Limiting cases for 2 alleles}
\label{sec:2Alleles}

Setting $K = 2$ in Eq.~(\ref{KAlleleKolmogEq}) gives 
\begin{eqnarray}
\lefteqn{\frac{\partial  f_{\bf Z}({\bf z}, s; {\bf z}_0)}{\partial s} = } \nonumber\\
		&&	\qquad \frac{\partial}{\partial z_1} \left\{ (r_{12}z_1 - r_{21} z_2) f_{\bf Z}({\bf z}, s; {\bf z}_0) \right\} 
		     + 		    \frac{\partial}{\partial z_2} \left\{ (r_{21}z_2 - r_{12} z_1) f_{\bf Z}({\bf z}, s; {\bf z}_0) \right\}  \nonumber \\
		&&		\qquad\qquad       + \frac{e^{-s}}{\kappa_0} \left[ \frac{\partial^2}{\partial z_1^2} \{z_1 f_{\bf Z}({\bf z}, s; {\bf z}_0)\} 
			                                                                                   + \frac{\partial^2}{\partial z_2^2} \{z_2 f_{\bf Z}({\bf z}, s; {\bf z}_0)\} \right],
   \label{2AlleleKolmogEq}
\end{eqnarray}
while Bartlett's equation, Eq.~(\ref{BartlettKAllele}) becomes 
\begin{multline}
\frac{\partial \phi(\pmb{\theta}, s; {\bf z}_0)}{\partial s} + 
	\left\{ r_{12}(\theta_1 - \theta_2) + e^{-s} \theta_1^2 \right\} \frac{\partial \phi(\pmb{\theta}, s; {\bf z}_0)}{\partial \theta_1} \\ +
	\left\{ r_{21}(\theta_2 - \theta_1) + e^{-s} \theta_2^2 \right\} \frac{\partial \phi(\pmb{\theta}, s; {\bf z}_0)}{\partial \theta_2}   = 0.  
	   \label{Bartlett2Allele}
\end{multline}

\subsection{Stationary distribution}
\label{sec:Stationary}

We seek the long-term stationary distribution $f_{\bf Z}({\bf z}, \infty; {\bf z}_0)$ of the scaled population corresponding to the initial condition 
$f_{\bf Z}({\bf z}, 0; {\bf z}_0) = \delta({\bf z} - {\bf z}_0)$.  The derivation given here is heuristic rather than rigorous, and will be tested for 
consistency with numerical simulations in Section~\ref{sec:NumericalSimulations}.  
We will assume without proof that Eq.~(\ref{Bartlett2Allele}) with initial condition 
$\phi(\pmb{\theta}, 0; {\bf z}_0) = e^{-\kappa_0(\theta_1 z_1 + \theta_2 z_2)}$ has a unique solution with a stationary, stable limit as $s \rightarrow \infty$ 
whose inverse Laplace transform is $f_{\bf Z}({\bf z}, \infty; {\bf z}_0)$.  

Under the assumption that the terms 
$\partial \phi/\partial s$, $e^{-s} \theta_1^2 \partial \phi/\partial \theta_1$  and $e^{-s} \theta_2^2 \partial \phi/\partial \theta_2$ become arbitrarily small  
in Eq.~(\ref{Bartlett2Allele}) for large $s$, the remaining terms dominate, giving
\begin{equation}
r_{12} \frac{\partial \phi(\pmb{\theta}, \infty; {\bf z}_0)}{\partial \theta_1} = r_{21} \frac{\partial \phi(\pmb{\theta}, \infty; {\bf z}_0)}{\partial \theta_2}.   
\end{equation}
The general solution to this partial differential equation is 
\begin{equation} 
\phi(\pmb{\theta}, \infty; {\bf z}_0) = g(r_{21}\theta_1 + r_{12}\theta_2), 
\end{equation}
where $g$ is an arbitrary function which is yet to be determined.  

Setting $\theta_1 = \theta_2 = \theta$, and using Eqs.~(\ref{laplaceKDim}), (\ref{laplace1Dim}) and the fact that $Z_{\rm tot}(s) = Z_1(s) + Z_2(s)$ is a 1-allele 
GW process independent of initial allele frequencies ${\bf z}_0$, it follows that 
\begin{eqnarray}
g((r_{12} + r_{21})\theta) &=& \phi((\theta, \theta), \infty; {\bf z}_0) \nonumber \\
	&=& E\left( e^{-\kappa_0 \theta Z_{\rm tot}(\infty)} | {\bf Z}(0) = {\bf z}_0 \right) \nonumber \\
	&=& E\left( e^{-\kappa_0 \theta Z_{\rm tot}(\infty)} | Z_{\rm tot}(0) = 1 \right) \nonumber \\
	&=& \phi_\text{1-allele}(\theta, \infty; 1),  		\label{gOnDiagonal}
\end{eqnarray}
where the explicit form of $\phi_\text{1-allele}(\cdot)$ is given by Eq.~(\ref{Bartlett1AlleleSoln}). 
This determines the functional form of $g$ and therefore the solution 
\begin{equation}
\phi(\pmb{\theta}, \infty; {\bf z}_0) = \phi_\text{1-allele}\left(\frac{r_{21}\theta_1 + r_{12}\theta_2}{r_{12} + r_{21}}, \infty; 1\right).	\label{BartlettSolnAsymptotic}
\end{equation}
It can readily be verified by direct substitution into Eq.~(\ref{laplaceKDim}) that the inverse Laplace transform is 
\begin{eqnarray}
f_{\bf Z}({\bf z}, \infty; {\bf z}_0) & = & (r_{12} + r_{21}) \delta(r_{12}z_1 - r_{21}z_2) f_\text{1-allele}\left( z_1 + z_2, \infty; 1, \kappa_0 \right) \nonumber \\
	& = & (r_{12} + r_{21}) \delta(r_{12}z_1 - r_{21}z_2)f_\text{Feller}(z_1 + z_2; \kappa_0), 		\label{asymptoticSoln}
\end{eqnarray}
where $\delta$ is the Dirac delta function.
Numerical simulations presented in Section~\ref{sec:NumericalSimulations} below are consistent with this solution provided at least one of $r_{12}$ and $r_{21}$
 are strictly positive.  

The most interesting aspect of this solution is that the distribution collapses onto the line $r_{12}z_1 - r_{21}z_2 = 0$.  In other words, 
conditional on the population not becoming extinct, the ratio $Z_1/Z_2$ converges almost surely to $r_{21}/r_{12}$, independent of the initial $A_1$ 
abundance $z_{01}$.  In fact, the result is the continuum version of a particular case of the limit theorem quoted in Eq.~(\ref{limitTheorem}), 
with $\rho = \lambda$ and $\pmb{\nu} \propto (r_{21}\, r_{12})$.  
The physical import of this result is that the population partitions into two sub-populations, of types $A_1$ and $A_2$ respectively, in a ratio 
determined by the mutation rates.  
This is at variance with the case when mutation rates are set to zero, in which case the ratio $Z_1/Z_2$ maintains a distribution 
centred on the ratio $z_{01}/z_{20}$, the distribution being broad for $\kappa_0 < 1$ and narrow for $\kappa_0 > 1$~\citep{burden2016genetic}.  

Note that the behaviour has some similarity with the traditional 2-allele WF model with mutations, whose asymptotic distribution is 
well known to be a beta distribution~\citep{wright1931evolution}, which collapses onto a point mass for large population sizes: 
\begin{eqnarray}
f_{\rm WF}(x) & = & \frac{x^{2Mu_{21} - 1}(1 - x)^{2Mu_{12} - 1}}{B(2Mu_{12}, 2Mu_{21})}	\nonumber \\
	& \rightarrow & \delta \left(x - \frac{u_{21}}{u_{12} + u_{21}} \right) \quad \mbox{as } M \rightarrow \infty, \label{WFbeta}
\end{eqnarray}
for a population $M$ and fixed per-generation mutation rates $u_{12}$ and $u_{21}$.  Here $x$ is the proportion of the population carrying the $A_1$ allele.  
To compare with the asymptotic GW case, make the transformation 
$z_1 = x z_\text{tot}$, $z_2 = (1 - x) z_{\rm tot}$ in Eq.~(\ref{asymptoticSoln}) and return to the per-generation mutation rates via Eq.~(\ref{rijDef}) 
to obtain the corresponding density 
\begin{equation}
f_{X, Z_\text{tot}}(x, z_\text{tot}, \infty; {\bf z}_0) = \delta \left(x - \frac{u_{21}}{u_{12} + u_{21}} \right) f_\text{Feller}(z_\text{tot}; \kappa_0).
\end{equation}
The marginal distribution of $X$ clearly agrees the $M \rightarrow \infty$ WF limit and the marginal distribution of $Z_\text{tot}$ agrees with the $s \rightarrow \infty$ 
limit of Eq.~(\ref{ZTotDistr}).  However, it will become apparent in the next section that the transition to the asymptotic solution Eq.~(\ref{asymptoticSoln}) 
for finite $s$ is not via a beta distribution in the $x$ variable.  

\subsection{Low mutation rates}
\label{sec:LowMutationRates}

We can understand how the transition to two sub-populations occurs by studying the physically realistic case of low mutation rates.  For simplicity, we consider the 
case of equal mutations rates, and set  $r_{12} = r_{21} = r <<1$.  For $K = 2$ alleles, the characteristic equations Eq.~(\ref{characteristicKAlleles}) are 
\begin{equation}
\begin{split}
\frac{d\theta_1}{ds} = r(\theta_1 - \theta_2) + e^{-s} \theta_1^2, \\
\frac{d\theta_2}{ds} = r(\theta_2 - \theta_1) + e^{-s} \theta_2^2. 
\end{split}		\label{characteristic2Alleles}
\end{equation}

\begin{figure}[t!]
\begin{center}
\centerline{\includegraphics[width=0.75\textwidth]{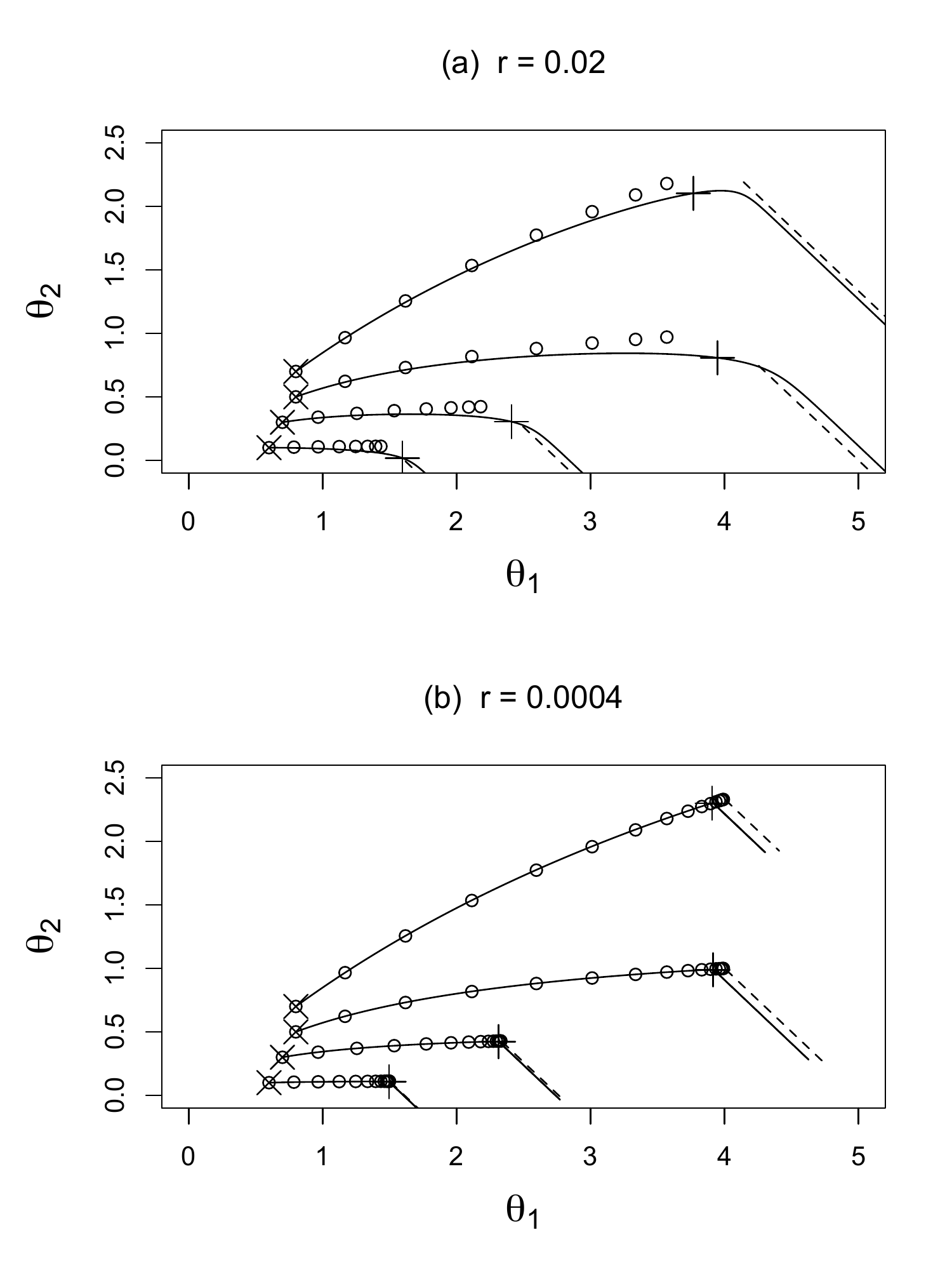}}
\caption{
Plots of trajectories in the Laplace-transformed $\theta_1$-$\theta_2$ plane corresponding to 
numerical solutions to the characteristic equations for 2 alleles, 
Eq.~(\ref{characteristic2Alleles}), (solid curve) for (a) $r = 0.02$ and (b) $r = 0.0004$ 
for several initial coordinates $(\theta_1(0), \theta_2(0))$ indicated as $\times$.  The changeover points Eq.~(\ref{changeover2Alleles}) are indicated as $+$.  
The circles are the approximate solution for $s < -\log r$, Eq.~(\ref{char2AllelePhase1}), and the dashed lines are the approximate solution for $s > -\log r$, Eq.~(\ref{char2AllelePhase2}).  These values of $r$ match those used in numerical simulations in Section~\ref{sec:NumericalSimulations}.  }
\label{fig:characteristic2Alleles}
\end{center}
\end{figure}

Numerical solutions to these equations in Figure~\ref{fig:characteristic2Alleles} show that if $0 < r << 1$ the characteristics undergo a rapid change in 
behaviour at $s = -\log r$ determined by whether the first or second term on the right hand side of each equation dominates.  More specifically, 
\begin{equation}
\begin{split}
\frac{d\theta_1}{ds} \approx \begin{cases}
e^{-s} \theta_1^2, & \mbox{if } s < -\log r, \\
r(\theta_1 - \theta_2), & \mbox{if } s > -\log r, 
\end{cases} \\ 
\frac{d\theta_2}{ds} \approx \begin{cases}
e^{-s} \theta_2^2, & \mbox{if } s < -\log r, \\
r(\theta_2 - \theta_1), & \mbox{if } s > -\log r. 
\end{cases}   
\end{split}		\label{characteristicSplit}
\end{equation}
Tracing back the origins of these terms to the forward Kolmogorov equation, Eq.~(\ref{2AlleleKolmogEq}), we observe that the earlier phase, 
$s < -\log r$, is dominated by genetic drift, and the later phase, $s > -\log r$, is dominated by mutations as the effect of genetic drift is diluted by exponential 
population growth.  
For $s < -\log r$ the approximate solution is Eq.~(\ref{characteristic1Allele}), that is  
\begin{equation}
\theta_1(s) \approx \frac{\theta_1(0)}{1 - (1 - e^{-s}) \theta_1(0)}, \qquad
\theta_2(s) \approx \frac{\theta_2(0)}{1 - (1 - e^{-s}) \theta_2(0)}, \qquad (s < -\log r),	\label{char2AllelePhase1}
\end{equation}
so $\theta_1(s)$ and $\theta_2(s)$ are simply two copies of the 1-allele case.  
The corresponding solution for the density $f_{\bf Z}$ for an initial condition
\begin{equation}
f_{\bf Z}({\bf z}, 0; {\bf z}_0) = \delta(z_1 - z_{01}) \delta(z_2 - 1 + z_{01}),  \label{initCond2Alleles}
\end{equation}
is, from Eq.~(\ref{zeroMutationDensity}), 
\begin{equation}
f_{\bf Z}({\bf z}, s; {\bf z}_0) \approx f_\text{1-allele}(z_1, s; z_{01}, \kappa_0) f_\text{1-allele}(z_2, s; 1 - z_{01}, \kappa_0),  \qquad (s < -\log r), 	\label{soln2AllelePhase1}
\end{equation} 
where the function $f_\text{1-allele}$ is defined in Eq.~(\ref{1AlleleSoln}).  
As it stands this approximate solution for $s < -\log r$ is too crude an approximation to be of use in analysing the biological data in the usual situation in which 
SNPs are rare within the genome. Essentially it tells us that, if we start with a non-segregating site for which $z_{01} = 0$ at $s = 0$, then the site is unlikely to 
manifest as a SNP provided $s < -\log r$.  

The more interesting case occurs after the changeover point, when $s > -\log r$.  The changeover point in the $(\theta_1, \theta_2)$ plane has coordinates 
determined from Eq.~(\ref{char2AllelePhase1}), 
\begin{equation}
\left( \begin{array}{c} 
c_1 \\ c_2 
\end{array} \right)
= 
\left( \begin{array}{c} 
\theta_1(-\log r) \\ \theta_2(-\log r) 
\end{array} \right) 
= 
\left( \begin{array}{c} 
\theta_1(0)/(1 - \theta_1(0)) \\ \theta_2(0)/(1 - \theta_2(0)) 
\end{array} \right) + O(r).	\label{changeover2Alleles}
\end{equation}

For $s > -\log r$ the general solution to the approximate characteristic equations is 
\begin{equation}
\left( \begin{array}{c} 
\theta_1(s) \\ \theta_2(s) 
\end{array} \right)
\approx 
a \left( \begin{array}{c} 
1 \\ 1 
\end{array} \right)
+ b \left( \begin{array}{c} 
1 \\ -1 
\end{array} \right) e^{2rs}.  \label{char2AlleleGen}
\end{equation}
The arbitrary constants $a$ and $b$ are determined from the coordinates of the changeover point, 
 \begin{equation}
\left( \begin{array}{c} 
c_1 \\ c_2 
\end{array} \right)
= 
a \left( \begin{array}{c} 
1 \\ 1 
\end{array} \right)
+ b \left( \begin{array}{c} 
1 \\ -1 
\end{array} \right) e^{-2r\log r}
=   
\left( \begin{array}{c} 
a + b \\ a - b 
\end{array} \right) + O(r\log r), 
\end{equation}
giving $a \approx (c_1 + c_2)/2$ and $b \approx (c_1 - c_2)/2$, assuming $|r\log r| << 1$.  Substituting back into Eq.~(\ref{char2AlleleGen}) gives 
\begin{equation}
\begin{split}
\theta_1(s) &\approx \tfrac{1}{2}(1 + e^{2rs}) c_1 + \tfrac{1}{2}(1 - e^{2rs}) c_2, \\
\theta_2(s) &\approx \tfrac{1}{2}(1 - e^{2rs}) c_1 + \tfrac{1}{2}(1 + e^{2rs}) c_2 , \qquad (s > -\log r),	\label{char2AllelePhase2}
\end{split}
\end{equation}
where 
\begin{equation}
c_1 = \frac{\theta_1(0)}{1 - \theta_1(0)}, \qquad c_2 = \frac{\theta_2(0)}{1 - \theta_2(0)}.     \label{CDef}
\end{equation}
Examples of characteristic curves with the approximate characteristics, Eqs.~(\ref{char2AllelePhase1}) and (\ref{char2AllelePhase2}) 
superimposed are plotted in Fig.~\ref{fig:characteristic2Alleles}.   

To solve Eq.~(\ref{Bartlett2Allele}), it is necessary to locate the initial 
coordinate $(\theta_1(0), \theta_2(0))$ in terms of a given final coordinate $(\theta_1(s), \theta_2(s))$.  From Eq.~(\ref{char2AllelePhase2}) the coordinates of the 
changeover point traced back from $(\theta_1(s), \theta_2(s))$ are 
\begin{equation}
c_1 = \alpha(s) \theta_1(s) + \beta(s) \theta_2(s), \qquad c_2 = \beta(s) \theta_1(s) + \alpha(s) \theta_2(s),  
\end{equation}
where
\begin{equation}
\alpha(s) = \tfrac{1}{2}(1 + e^{-2rs}), \qquad \beta(s) = \tfrac{1}{2}(1 - e^{-2rs}). \label{alphaBetaDef}
\end{equation}
Then from Eq.~(\ref{CDef}), the initial coordinate is 
\begin{equation}
\begin{split}
\theta_1(0) = \frac{c_1}{1 + c_1} = \frac{ \alpha(s) \theta_1(s) + \beta(s) \theta_2(s)}{1 +  \alpha(s) \theta_1(s) + \beta(s) \theta_2(s)}, \\
\theta_2(0) = \frac{c_2}{1 + c_2} = \frac{ \beta(s) \theta_1(s) + \alpha(s) \theta_2(s)}{1 +  \beta(s) \theta_1(s) + \alpha(s) \theta_2(s)}.  
\end{split}	\label{initCharacteristicCoord}
\end{equation}
The value of $\phi$ on the boundary at $s = 0$ is, from Eqs.~(\ref{laplaceKDim}) and (\ref{initCond2Alleles}), 
\begin{equation}
\phi(\pmb{\theta}, 0; {\bf z}_0)  = e^{-\kappa_0 \{\theta_1(0) z_{01} + \theta_2(0)(1 - z_{01})\}}.  \label{initPhi2Alleles}
\end{equation}
Combining Eqs.~(\ref{initCharacteristicCoord}) and (\ref{initPhi2Alleles}), and the principle that $\phi$ is constant along the characteristics gives the 
approximate solution to Bartlett's equation\footnote{We note in passing that the limit as $s \rightarrow \infty$ of Eq.~(\ref{BartlettSolnSmallR}) agrees with Eq.~(\ref{BartlettSolnAsymptotic}) with $r_{12}$ set equal to $r_{21}$.  This is required to be the case since Eq.~(\ref{BartlettSolnAsymptotic}) 
is the solution to the 2-allele Bartlett's equation for arbitrary $r_{12}$ and $r_{21}$ in the limit $s \rightarrow \infty$, whereas Eq.~(\ref{BartlettSolnSmallR}) is the solution 
for arbitrary $s > -\log r$ provided $r_{12} = r_{21} <<1$.}  
\begin{equation}
\phi(\pmb{\theta}, s; {\bf z}_0) \approx \phi_1(\pmb{\theta}, s; {\bf z}_0) \phi_2(\pmb{\theta}, s; {\bf z}_0), 	\label{BartlettSolnSmallR} 
\end{equation}
where 
\begin{equation}
\begin{split}
\phi_1(\pmb{\theta}, s; {\bf z}_0) &= \exp\left\{-\kappa_0 z_{01} \frac{ \alpha(s) \theta_1 + \beta(s) \theta_2}{1 +  \alpha(s) \theta_1 + \beta(s) \theta_2} \right\} \\
		&= \phi_\text{1-allele}( \alpha(s) \theta_1 + \beta(s) \theta_2, \infty; z_{01}), \\
\phi_2(\pmb{\theta}, s; {\bf z}_0) &= \exp\left\{-\kappa_0 (1 - z_{01}) \frac{ \beta(s) \theta_1 + \alpha(s) \theta_2}{1 +  \beta(s) \theta_1 + \alpha(s) \theta_2} \right\} \\ 
		&= \phi_\text{1-allele}( \beta(s) \theta_1 + \alpha(s) \theta_2, \infty; 1 - z_{01}), 
\end{split}
\end{equation}
where the function $\phi_\text{1-allele}$ is defined by Eq.~(\ref{Bartlett1AlleleSoln}).

The inverse of the Laplace transform of Eq.~(\ref{BartlettSolnSmallR}) is a convolution integral, 
\begin{equation}
f_{\bf Z}({\bf z}, s; {\bf z}_0) = \int_{\mathds{R}^2} d^2{\bf u} \, f_1({\bf u}, s; {\bf z}_0) f_2({\bf z} - {\bf u}, s; {\bf z}_0), \label{convIntegral}
\end{equation}
where $f_1$ and $f_2$ are the inverse Laplace transforms of $\phi_1$ and $\phi_2$ respectively.  It is straightforward to check by substitution into 
Eq.~(\ref{laplaceKDim}) and a change of variables that these inverse Laplace transforms are 
\begin{equation}
\begin{split}
f_1({\bf z}, s; {\bf z}_0) = f_\text{1-allele}\left(\frac{\alpha(s) z_1 + \beta(s) z_2}{\alpha^2 + \beta^2}, \infty; z_{01}, \kappa_0 \right) \delta(\alpha(s) z_2 - \beta(s) z_1), \\
f_2({\bf z}, s; {\bf z}_0) = f_\text{1-allele}\left(\frac{\beta(s) z_1 + \alpha(s) z_2}{\beta^2 + \alpha^2}, \infty; 1 - z_{01}, \kappa_0 \right) \delta(\beta(s) z_2 - \alpha(s) z_1), 
\end{split}
\end{equation}
where the inverse Laplace transform of $\phi_\text{1-allele}$, namely $f_\text{1-allele}$, is given in Eq.~(\ref{1AlleleSoln}).  The delta functions enable the 
integral in Eq.~(\ref{convIntegral}) to be carried through.  The final result is 
\begin{equation}
\begin{split}
f_{\bf Z}({\bf z}, s; {\bf z}_0) & \approx  \frac{1}{\alpha(s)^2 - \beta(s)^2} \times   \\
&		f_\text{1-allele}\left(\frac{\alpha(s) z_1 - \beta(s) z_2}{\alpha(s)^2 - \beta(s)^2}, \infty; z_{01}, \kappa_0 \right)  \times  \\
&		f_\text{1-allele}\left(\frac{\alpha(s) z_2 - \beta(s) z_1}{\alpha(s)^2 - \beta(s)^2}, \infty; 1 - z_{01}, \kappa_0 \right), 		  \qquad (s > -\log r). 
\end{split}		\label{soln2AllelePhase2}
\end{equation}
The interesting aspect of this distribution is that, since $f_\text{1-allele}$ is only non-zero for non-negative arguments, $f_{\bf Z}$ is only non-zero 
if both $\alpha(s) z_1 - \beta(s) z_2 \ge 0$ and $\alpha(s) z_2 - \beta(s) z_1 \ge 0$.  From Eqs.~(\ref{alphaBetaDef}), it follows that 
\begin{equation}
\tanh rs = \frac{\beta(s)}{\alpha(s)} \le \frac{Z_2(s)}{Z_1(s)} \le \frac{\alpha(s)}{\beta(s)} = \coth rs.  \label{boundaryApproxSoln}
\end{equation}
That is to say, the support of the distribution is sandwiched between lines in the $z_1$-$z_2$ plane of slope $\coth rs$ and $\tanh rs$.  
Equivalently,  the proportion of the population carrying the $A_1$ allele, namely $X = Z_1/(Z_1 + Z_2)$, is sandwiched between the 
values $\frac{1}{2}(1 \pm e^{-2rs})$, with a high density of probability at the end points due to the delta function in $f_\text{1-allele}$.  
As $s \rightarrow \infty$ the distribution converges on a line with slope 1, consistent with the stationary distribution found in Section~\ref{sec:Stationary}.  
Note however that for finite $s$ the marginal distribution of $X$ differs from the WF beta distribution of Eq.~(\ref{WFbeta}), whose support for finite $M$ 
is the entire interval $[0, 1]$.  


\section{Numerical Simulations}
\label{sec:NumericalSimulations}

\begin{figure}[t!]
\begin{center}
\centerline{\includegraphics[width=0.88\textwidth]{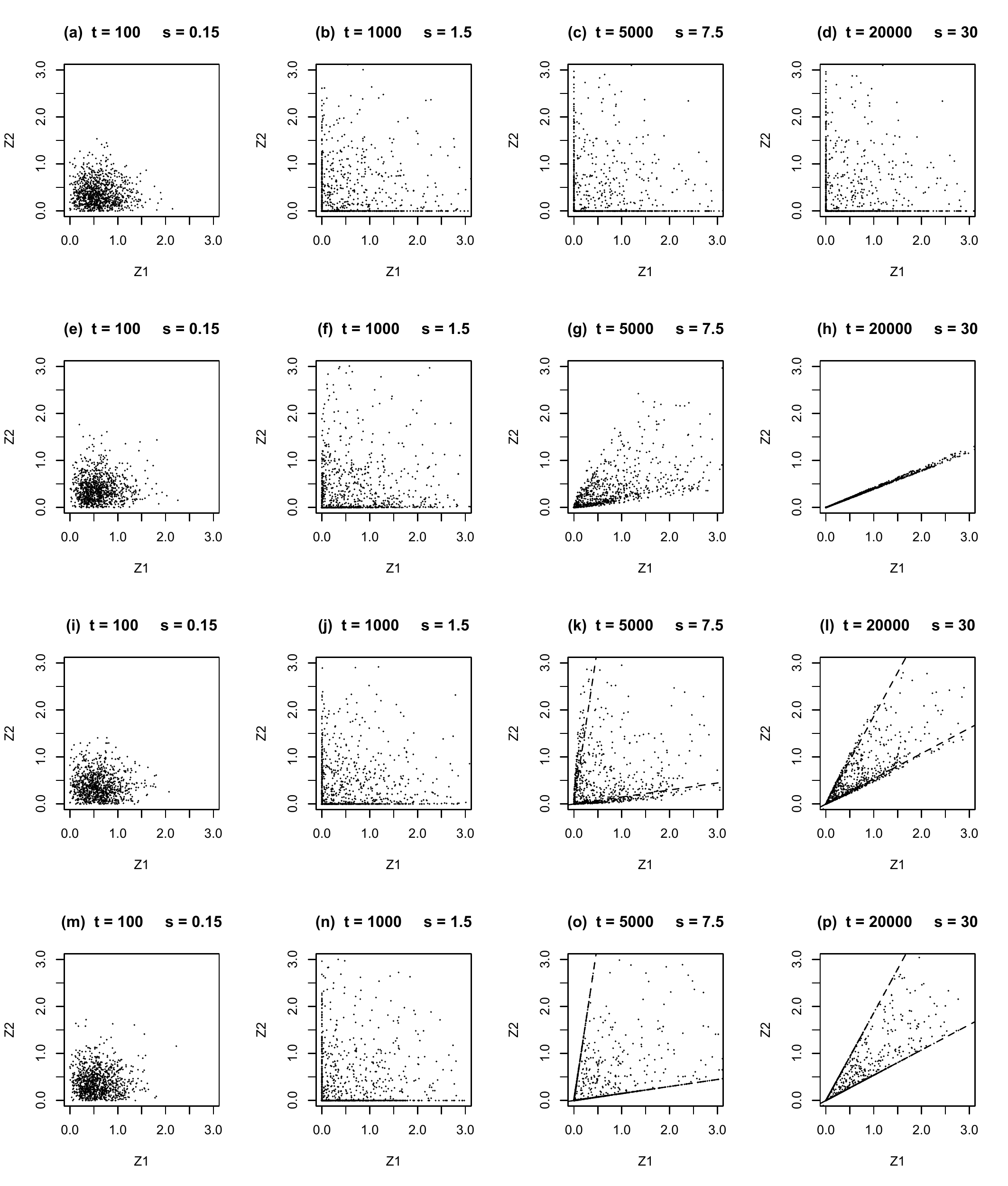}}
\caption{Numerical simulations of the  of the multitype branching model of neutral evolution.  Plots show the scaled allele abundances $Z_1(s)$ and 
$Z_2(s)$ defined by Eq.~(\ref{ZDef}).  Parameter values in all simulations are $m_0 = 1000$, $\log \lambda = 0.0015$, $\sigma^2 = 2$, and $z_{01} = 0.6$. 
Mutation rates are: (a) to (d), $u_{12} = u_{21} = 0$; (e) to (h) $u_{12} = 6 \times 10^{-5}$ and $u_{21} = 1.5 \times 10^{-4}$; 
(i) to (p), $u_{12} = u_{21} = 3 \times 10^{-5}$.  Plots (a) to (l) are a simulation of the model described in Section~\ref{sec:TheModel}.  
Plots (m) to (p) are generated from the theoretical distribution, Eqs.~(\ref{soln2AllelePhase1}) and (\ref{soln2AllelePhase2}). } 
\label{fig:ScatterPlot}
\end{center}
\end{figure}

We have carried out several numerical simulations of the multitype branching model of neutral evolution described in Section~\ref{sec:TheModel}  
for the case of $K = 2$ allele types.  Each simulation shown in Fig.~\ref{fig:ScatterPlot} begins with an initial population of $m_0 = 1000$ individuals, 
of whom $600$ are of allele type 1 and $400$ are of allele type 2.  The number of offspring produced by 
any individual in any generation is an i.i.d.\ negative binomial random variable with mean $\lambda$ such that $\log \lambda = 0.0015$ and with variance 
$\sigma^2 = 2$.  The corresponding scaled parameters introduced in Section~\ref{sec:DiffusionLimit} and used throughout the subsequent simulations are 
\begin{equation}
\kappa_0 = 1.5, \qquad z_{01} = 0.6, \quad z_{02} = 0.4.  
\end{equation}
The initial population size and growth rate are chosen to mimic a simulation of the female part of the human population during the upper Paleolithic 
period carried out in~\citet{burden2016genetic}, beginning from the time of mitochondrial Eve (mtE) and ending at the boundary between the Paleolithic and 
Neolithic periods.  This corresponds to approximately $t = 5000$ to $6000$ generations, or a scaled time period of $s = t \log \lambda$ in the range 7.9 
to 9.  However, to illustrate the asymptotic behaviour, the simulations were continued to $20000$ generations.  At each set of parameter values 1000 
trajectories were computed. Scaled allele abundances $Z_1(s)$ and $Z_2(s)$ at various time points and for various choices of mutation rates are plotted in 
Fig.~\ref{fig:ScatterPlot}.  

In plots (a) to (d) the per-generation mutation rates $u_{12}$ and $u_{21}$ are set to zero.  In this case the two 
alleles evolve independently according to Eq.~(\ref{zeroMutationDensity}).  By $t = 5000$ generations, or $s = 7.5$, the factor $(1 - e^{-s})^{-1}$ in 
Eq.~(\ref{kappaTildeDef}) is close to 1, and the distribution is indistinguishable from its asymptotic form.  
This scenario is discussed in detail by~\citet{burden2016genetic}.  
 
In plots (e) to (h) we have set the per-generation mutation rates to $u_{12} = 6 \times 10^{-5}$ and $u_{21} = 1.5 \times 10^{-4}$.  The corresponding 
scaled mutation rates, defined by Eq.~(\ref{rijDef}) are $r_{12} = 0.04$ and $r_{21} = 0.1$.  Consistent with the results of Section~\ref{sec:Stationary}, 
the distribution converges on the line $Z_1/Z_2 = r_{21}/r_{12} = 2.5$ as $t \rightarrow \infty$.  

In plots (i) to (l) the per-generation mutation rates is set to to $u_{12} = u_{21} = 3 \times 10^{-5}$, corresponding to  $r_{12} = r_{21} = 0.02$.  
For comparison, plots (m) to (p) are generated randomly from the theoretical distribution valid for $r_{12} = r_{21} = r << 1$ determined in 
Section~\ref{sec:LowMutationRates}.  Recall the theoretical prediction that the distribution undergoes a rapid changeover from Eq.~(\ref{soln2AllelePhase1}) 
for $s < -\log r$ to Eq.~(\ref{soln2AllelePhase2}) for $s > -\log r$.  For these parameters the changeover point occurs at $s \approx 3.91$ or $t \approx  2608$, 
that is, between the second and third columns of Fig.~(\ref{fig:ScatterPlot}).  Since Eq.~(\ref{soln2AllelePhase1}) corresponds to two independent GW 
branching processes, plots (m) and (n) are easily generated by sampling $Z_1$ and $Z_2$ independently from Feller's solution, Eq.~(\ref{fellerDistrDef}), 
with appropriate scaling.  As expected, plots (m) and (n) are consistent not only with plots (i) and (j), but also with plots (a) and (b), for which the mutation 
rate is set to zero.  

To generate plots (o) and (p), we first defined random variables 
\begin{equation}
U_1 = \frac{1}{z_{01}}\frac{\alpha(s) Z_1 - \beta(s) Z_2}{\alpha(s)^2 - \beta(s)^2},  \quad 
U_2 = \frac{1}{1 - z_{01}}\frac{\alpha(s) Z_2 - \beta(s) Z_1}{\alpha(s)^2 - \beta(s)^2},  \label{ZToUTransf}
\end{equation}
whose joint distribution density is found from Eqs.~(\ref{soln2AllelePhase2}) and (\ref{f1AlleleAndFeller}) to be 
\begin{eqnarray}
f_{\bf U}({\bf u}, s, {\bf z}_0) & = & f_{\bf Z}({\bf z}, s, {\bf z}_0) \left| \frac{\partial (z_1, z_2)}{\partial (u_1, u_2)} \right| \nonumber \\
	& = & f_\text{Feller}(u_1, \kappa_0 z_0) f_\text{Feller}(u_2, \kappa_0 (1 - z_0)).  
\end{eqnarray}
We obtained the required sample by first sampling $U_1$ and $U_2$ from Feller's solution, and then transforming via the inverse of 
Eq.~(\ref{ZToUTransf}), namely 
\begin{equation}
Z_1 = \alpha(s) z_0 U_1 + \beta(s) (1 - z_0) U_2, \quad Z_2 = \beta(s) z_0 U_1 + \alpha(s) (1 - z_0) U_2. 
\end{equation}
The boundary of the support of the approximate distribution, Eq.~(\ref{boundaryApproxSoln}), is shown as a dashed line in plots (k), (l), (o) and (p).  
The simulations in plots (k) and (l) are clearly converging on to the diagonal $Z_1 = Z_2$ as $s \rightarrow \infty$ in reasonable agreement with 
the approximate theory.  

The mutation rates in the above simulations are chosen to illustrate the comparison between the multitype GW branching model and 
mathematical properties of solutions to the forward Kolmogorov equation.  However they are considerably higher than observed genomic 
mutation rates.  For instance, a recent survey by~\citet{kivisild2015maternal} quotes a mutation rate for synonymous sites in human mitochondrial DNA 
in the order of $3 \times 10^{-8}$ per base pair per year.  Translating this to the simulation of the Paleolithic human population 
mentioned at the beginning of this section
equates to a rate\footnote{For the purposes of the following simulation we interpret allele type 1 and allele type 2 to be 
any major allele and minor allele respectively at a given site.} $u_{12} = u_{21} = 6 \times 10^{-7}$ per 20 year generation, or a scaled rate 
$r_{12} = r_{21} = 4 \times 10^{-4}$.  This places the changeover point separating the behaviour described by Eq.~(\ref{soln2AllelePhase1}) from 
the behaviour described by Eq.~(\ref{soln2AllelePhase2}) at $s = -\log r \approx  7.82$, or approximately $t = 5200$ generations.  Coincidentally 
this is close to the estimated time of 5610 generations between mtE and the end of the Paleolithic obtained by~\citet{burden2016genetic}.  

Unfortunately the approximate solution found in Section~\ref{sec:LowMutationRates} is too crude to be of much help to us up to the changeover 
point as it simply tells us that the solution to the forward Kolmogorov equation closely approximates the model with no mutation.  Instead we have resorted to  
the following simulation of the multitype branching model, which we will compare with observations of segregating synonymous sites in 
mitochondrial DNA~\citep{kivisild2005role}.  The simulation is based on the assumed scenario that the female population 
a the end of the Paleolithic, $M(5610)\approx 3 \times 10^6$, is descended from a single individual ($m_0 = 1$) of specified allele type ($z_{01} = 1$), 
namely mtE.  The simulation consisted of 500000 runs of a multitype branching process with the above parameters.  Those with a final population falling 
outside the range $1.6 \times 10^6 <  M(5610) < 4.8 \times 10^6$ were discarded.  As expected, the vast majority of runs correspond to populations which drop to zero, that is, 
lineages which become extinct.  The number of runs surviving the filter, namely 527, is in accordance with expectations given that the estimate of the population 
at the time of mtE is of order $10^3$.  

\begin{figure}[t!]
\begin{center}
\centerline{\includegraphics[width=0.75\textwidth]{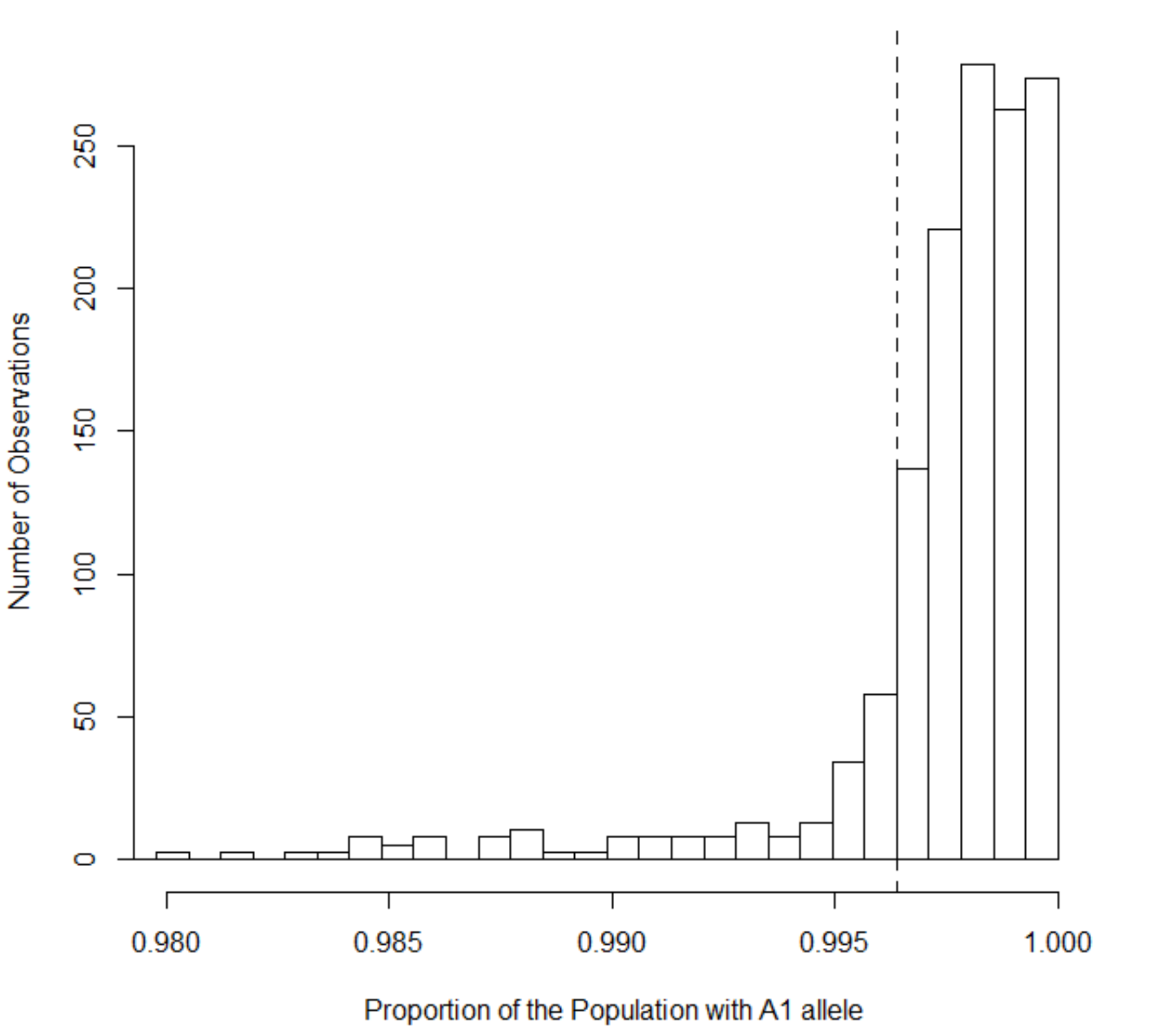}}
\caption{Histogram of the proportion $Y_1(t)/(Y_1(t) + Y_2(t))$ of the population with type-1 alleles at the end of the Paleolithic period $t = 5610$, from the 
527 valid runs out of 500000 simulations, each starting with an initial population of $m_0 = 1$ type-1 individual.  
The simulations were performed assuming the multitype branching process defined in Section~\ref{sec:TheModel}, assuming the number of offspring per 
individual per generation to be an i.i.d.\ negative binomial random variable.  
Parameters used are $\log \lambda = 0.0015$, $\sigma^2 = 2$, 
$u_{12} = u_{21} = 6 \times 10^{-7}$.  The dashed vertical line represents the threshold $1 - \frac{1}{277}$, below which a site is considered to be a SNP.}  
\label{fig:PaleolithicSim}
\end{center}
\end{figure}

Fig.~\ref{fig:PaleolithicSim} is a histogram of the proportion $Y_1(t)/(Y_1(t) + Y_2(t))$ at $t = 5610$ of type-1 alleles in the surviving 527 runs.  We interpret 
this histogram as a proxy for the site frequency spectrum of major alleles among neutrally evolving genomic sites.  Assuming that the site frequency spectrum 
has not changed markedly in the 12000 years since the end of the Paleolithic period, we compare our simulation with the empirical study of~\citet{kivisild2005role} 
who sampled a total of 277 individual human genomes.  Given the sample size, unless a SNP is prevalent in at least a fraction $\frac{1}{277}$ of the population, 
it is unlikely that it will be observed.  Therefore we set a threshold at $1 - \frac{1}{277}$, and classify any simulation whose fraction of type-1 alleles falls below 
this threshold as a SNP, while those falling above the threshold are classified as non-segregating sites.  A total of 79 out of 527 simulations (15.0\%) were found 
to yield a fraction of type-1 alleles below the threshold.  The result is in broad agreement with Table~1 of~\citet{kivisild2005role}, in which a total of 785 out of 4212 
synonymous sites (18.6\%) are observed to be segregating.  


\section{Discussion and conclusions}
\label{sec:Discussion}

The main focus of this paper is the diffusion limit of multitype branching processes, presented as a model of the evolution of genomic frequencies in a 
growing population.  In particular we have considered the case of neutral mutations between two alleles.  While the full forward Kolmogorov equation remains 
intractable, solutions are found in two limiting cases: the asymptotic stationary distribution at large times, and an approximate solution to the evolving allele 
frequency distribution for small scaled mutation rates.  

The asymptotic solution is  a manifestation of a well known result for multitype branching processes, 
encapsulated in Eq.~(\ref{limitTheorem}), namely that as $t \rightarrow \infty$ the population partitions almost surely into two subpopulations corresponding to the two alleles, in 
the ratio of the two mutation rates.  Also of interest is the path by which the population arrives at its asymptotic state,  
which brings us to the second, and more important result concerning the approximate solution in the biologically relevant limit of small mutation rates.  
We find that evolution of the allele distribution proceeds in two phases, a drift-dominated phase described by Eq.~(\ref{soln2AllelePhase1}) and a 
mutation-dominated phase described by Eq.~(\ref{soln2AllelePhase2}), 
separated by a changeover point at the scaled time $s_{\text c} = -\log r$.  In terms of the unscaled parameters, the changeover point, measured in generations 
after a given starting population, is 
\begin{equation}
t_{\text c} = \frac{-\log(u/\log\lambda)}{\log\lambda}, 
\end{equation}
where $u$ is the per-generation mutation rate (assumed equal in both directions) and $\lambda$ per-generation growth factor.  For $t < t_{\text c}$ the 
distribution is, not surprisingly, a perturbation on the case of zero mutation rates.  However for $t > t_{\text c}$ the solution changes dramatically in that the 
support of the distribution is sandwiched between 
(see Eq.~(\ref{boundaryApproxSoln}) and Fig.~\ref{fig:ScatterPlot}) 
\begin{equation}
\tanh ut \le \frac{Y_2(t)}{Y_1(t)} \le \coth ut. 
\end{equation}
Note that the mutation-dominated phase only occurs for for super-critical growth, since $t_{\text c} \rightarrow \infty$ as $\lambda \rightarrow 1_+$ for fixed $u$.

Significantly, if the starting population is entirely of allele type 1, then after $t_{\text c}$ generations the proportion of type-2 alleles will be bounded below.  
That is to say, at some level of sampling {\em every} neutral genomic site must eventually manifest as segregating with minor allele frequency bounded below by $\tanh ut$.  
However, the mutation rates used to illustrate the point in Fig.~\ref{fig:ScatterPlot} are orders of magnitude higher than observed mutation rates in most species.  By comparison, 
the simulation leading to Fig.~\ref{fig:PaleolithicSim} assumes mutation rates for neutral mitochondrial sites of $u = 6\times 10^{-7}$ per generation, 
and computes the major allele frequency distribution after 5200 generations at a time we associate with the Paleolithic/Neolithic transition.  This time is close 
to $t_{\text c}$, and we do not expect the lower bound to apply.  In order to see a lower bound of $0.005$ or $0.01$ respectively on the minor allele frequency one 
would need the Paleolithic growth rate to continue to $t = 8333$ or $16667$ generations, by which time the expected population would have grown to $2.7 \times 10^8$ 
or to an astronomical $7.2 \times 10^{13}$ respectively.  
Total collapse onto a partitioned population 
as seen in Fig.~\ref{fig:ScatterPlot}(h) would clearly require the 
population to grow well past any planet's carrying capacity.  If these growth and neutral mutation rates are typical of other populations of organisms~\citep{drake1998rates}, 
it seems unlikely that
this partitioning could occur via the above mechanism solely through neutral mutations.  Having said that, we note that one possible exception 
may be RNA viruses, for which mutation rates are orders of magnitude  higher than in DNA based organisms~\citep{lauring2013role}.   

It should be possible to extend the mathematical analysis in this paper to include selection, multiple alleles and arbitrary instantaneous rate matrices, 
all of which are implicit in the generic discrete-state model of Section~\ref{sec:TheModel}.  The limit theorem embodied in Eq.~(\ref{limitTheorem}) still applies, 
implying that the population will again partition into alleles as the asymptotic distribution collapses onto the direction of an eigenvector of the matrix 
in Eq.~(\ref{muDef}).  Perhaps of more immediate biological relevance, however, is an understanding of the nature of the solution in the period leading up to the changeover 
point, particularly how it compares with the analogous WF based coalescent models.  

%
%
\section*{Acknowledgements}

The authors wish to thank two anonymous reviewers for a number of helpful observations and suggestions.  

\section*{References}


\end{document}